\documentclass[aps, prd, superscriptaddress, nofootinbib, preprintnumbers]{revtex4}

\usepackage{amsmath,amssymb,bm}
\usepackage[dvipdfmx]{graphicx}
\usepackage{color}
\usepackage{float}
\usepackage{braket}

\allowdisplaybreaks

\begin{document}

\title{ 
Lifetime and mass of rho meson  in correlation with \\ 
magnetic-dimensional reduction 
}
\author{Mamiya Kawaguchi\footnote{mkawaguchi@hken.phys.nagoya-u.ac.jp}}
      \affiliation{ Department of Physics, Nagoya University, Nagoya 464-8602, Japan.} 
\author{Shinya Matsuzaki\footnote{synya@hken.phys.nagoya-u.ac.jp}}
      \affiliation{ Department of Physics, Nagoya University, Nagoya 464-8602, Japan.}       
\affiliation{ Institute for Advanced Research, Nagoya University, Nagoya 464-8602, Japan.}

\date{\today}

\def\theequation{\thesection.\arabic{equation}}
\makeatother

 \begin{abstract}
It is simply anticipated that    
in a strong magnetic configuration, 
the Landau quantization 
ceases the neutral rho meson to decay to the charged pion pair,  
so the neutral rho meson  will be long-lived. 
To closely access this naive observation, 
we explicitly compute  
the charged pion-loop in the magnetic field at the one-loop level, 
to evaluate the magnetic dependence of the lifetime for the neutral 
rho meson as well as its mass. 
Due to the dimensional reduction induced by the magnetic field 
(violation of the Lorentz invariance), 
the polarization (spin $s_z=0,\pm 1$) 
modes of the rho meson, as well as 
the corresponding pole mass and width, are decomposed 
in a nontrivial manner compared to the vacuum case. 
To see the significance of the reduction effect, 
we simply take the lowest-Landau level approximation 
to analyze the spin-dependent rho masses and widths.  
We find that the ``fate" of the rho meson may be more complicated 
 because of the magnetic-dimensional reduction:  
as the magnetic field increases,   
the rho width for the spin $s_z=0$ starts to develop, 
reach a peak, to be vanishing at the critical magnetic field 
to which the folklore refers. 
On the other side, the decay rates of 
the other rhos for $s_z=\pm 1$ monotonically increase 
as the magnetic field develops. 
The correlation between the polarization dependence and 
the Landau-level truncation is also addressed.  
\end{abstract}
\maketitle
%%%%%%%%%%%%%%%%%%%%%%%%%%%%%%%%%%%%%%%%
%%%%%%%%%%%%%%%%%%%%%%%%%%%%%%%%%%%%%%%%

\section{Introduction} 

Studies on hadron properties under the influence of 
the magnetic field have been extensively done so far. 
The development on such a hadron physics in the magnetic field would   
be relevant to gain some new insights for existing 
environment systems with the presence of strong magnetic fields, 
like in relativistic heavy ion collision~\cite{Kharzeev:2007jp,Skokov:2009qp}
and neutron stars~\cite{Harding:2006qn,Lattimer:2006xb,Potekhin:2011xe,Lai:2014nma}.

In the magnetic field, hadron properties will indeed be dramatically changed. 
For the neutral rho meson, in particular, it is expected that 
the decay channel to the charged pion pair, which is the main decay 
mode in the vacuum (without the magnetic configuration), will be closed, 
and hence the neutral rho meson can be long-lived. 
This is a naively-believed folklore, which can be reasoned by 
the Landau quantization for the charged pions: 
in the vacuum the neutral rho-meson decay width to the charged pion pair,  
$\Gamma(\rho\to \pi\pi)$, is given as 
\begin{equation} 
\Gamma(\rho^0\to \pi^+\pi^-)_{\rm vac}=
\frac{|g_{\rho\pi\pi}|^2}{6\pi m_\rho^2} 
\left(\sqrt{ \frac{m_\rho^2-4m_{\pi}^2}{4}} \right)^3
\,,  \label{width:vac}
\end{equation} 
with the $\rho^0-\pi^+-\pi^-$ coupling strength $g_{\rho \pi\pi}(\simeq 6)$. 
Now, turn on the magnetic field. 
Since the charged pions carry the electromagnetic 
charge, the mass ($m_{\pi}$) will be shifted by the magnetic field to be 
$\sqrt{m_{\pi}^2+eB}$ in the lowest Landau level (LLL). 
Then, one may naively suspect that 
the decay width in Eq.(\ref{width:vac}) will be 
modified in the phase space factor like    
\begin{equation}  
\Gamma(\rho^0\to \pi^+\pi^-)_{\rm naive} 
= \frac{|g_{\rho\pi\pi}|^2}{6\pi m_\rho^2} 
\left(\sqrt{\frac{ m_\rho^2-4(m_\pi^2+eB)}{4} } \right)^3
\,. \label{width:naive}
\end{equation} 
Thus, the $\rho^0 \to \pi^+ \pi^-$ decay channel is expected to be 
closed when the magnetic field reaches the critical scale $eB_c=(m_\rho^2 - 4 m_\pi^2)/4 
\simeq 0.13\,{\rm GeV}^2$.

Note that the above widely-accepted argument 
invokes only the kinematics, relying on the modification of 
the phase space factor:  
{\it the dynamical properties in the magnetic field, 
such as the dimensional reduction,   
are not taken into account at all}. 
Therefore, a more rigorous argument on this issue 
should involve some explicit dynamical computation of 
the rho width.

In this paper, 
we explicitly compute the charged pion-loop in the magnetic field at the one-loop level 
based on a chiral effective model.   
We evaluate the magnetic dependence of the decay width (lifetime) 
for the neutral rho meson as well as its mass. 
We find that the polarization (spin $s_z=0,\pm 1$) modes of the rho meson, as well as 
the corresponding pole mass and width, are decomposed 
in a nontrivial manner compared to the vacuum case. 
This is due to the dimensional reduction induced by the magnetic field.

In order to study the significance of the reduction effect, 
we simply take the LLL approximation 
and analyze the spin-dependent rho masses and widths.  
We find that as the magnetic field increases,   
the rho width for the spin $s_z=0$ starts to develop, 
reach a peak, to be vanishing at the critical magnetic field 
to which the folklore refers: 
this result would suggest that the ``fate" of the neutral rho meson 
in the magnetic field may be more complicated 
than the aforementioned naive expectation.  
On the other side, the decay rates of 
the other rhos for $s_z=\pm 1$ monotonically increase 
as the magnetic field develops. 
The correlation between the polarization dependence and 
the Landau-level truncation is also addressed.

This paper is constructed as follows: 
in Sec.~\ref{model} we introduce 
a chiral-effective model and 
extract the Lagrangian terms relevant to 
the later discussions. 
Sec.~\ref{loopcprrection} provides 
the explicit computation of the pion-loop contribution, 
including the effect of the magnetic-dimensional reduction,  
to the rho meson propagator at the one-loop level.  
In Sec.~\ref{numerical} 
we numerically evaluate the magnetic dependence of 
the neutral rho meson with respect to 
the polarization states intrinsically decomposed 
by the presence of the magnetic field. 
 Some possible interpretations for the results 
 obtained in Sec.~\ref{numerical} are proposed 
 in Sec.~\ref{dis}. 
 Summary is given in Sec.~\ref{summ}. 
The appendix~\ref{para-int} provides 
formulae for the Feynman parameter integrals 
relevant to evaluation of the one-loop terms given 
in Sec.~\ref{loopcprrection}.

%%%%%%%%%%%%%%%%%%%%%%%%%%%%%%%%%%%%%%%%
%%%%%%%%%%%%%%%%%%%%%%%%%%%%%%%%%%%%%%%%

\section{A chiral effective model} 
\label{model}

%%%%%%%%%%%%%%%%%%%%%%%%%%%%%%%%%%%%%%%%
%%%%%%%%%%%%%%%%%%%%%%%%%%%%%%%%%%%%%%%%
We employ a chiral effective model based 
on the coset space, 
$G/H=
[SU(2)_L\times SU(2)_R\times U(1)_V]/SU(2)_{V=L=R}\times U(1)_V$.  
The fundamental dynamical variables to construct  
the chiral effective Lagrangian are the nonlinear bases 
$\xi_{L,R}$, which transform 
under the $G$ as 
$\xi_{L,R}\rightarrow h(\pi,g_L,g_R)\cdot\xi_{L,R}\cdot g^\dagger_{L,R}$, 
where $h(\pi,g_L,g_R)\in H$ and $g_{L,R} \in G$. 
These variables are parameterized by the pions  
as $\xi_{L,R}=e^{\mp i\pi^a T^a/F_\pi}$ with $\pi^a$ $(a=1,2,3)$ being 
the pion fields, $T^a$ generators of $SU(2)$ normalized by 
${\rm tr}[T^a T^b]=\delta^{ab}/2$ 
and
$F_\pi$ is the pion decay constant. 
It is convenient to introduce the Maurer-Cartan 1-forms:
\begin{eqnarray}
\alpha_{\perp,||\mu}=\frac{1}{2i}(D_\mu\xi_R\cdot\xi^\dagger_R\mp D_\mu\xi_L\cdot\xi^\dagger_L) 
\,. \label{para-perp}
\end{eqnarray}
Here we have gauged the chiral symmetry with the external gauge 
fields ${\cal L}_{\mu}$ and ${\cal R}_\mu$, and  
\begin{eqnarray}
D_{\mu}\xi_{L}&=&\partial_\mu \xi_{L}+i\xi_{L}{\cal L}_\mu\,,\nonumber\\
D_{\mu}\xi_{R}&=&\partial_\mu \xi_{R}+i\xi_{R}{\cal R}_\mu\,,
\end{eqnarray} 
with the photon field $A_\mu$ incorporated as 
${\cal L}_\mu={\cal R}_\mu=eQ_{\rm em}A_\mu$ involving 
the electromagnetic coupling $e$ 
and the charge matrix 
$Q_{\rm em}=T^3+1/6 \cdot {\bm 1}_{2 \times 2} 
={\rm diag}(2/3,-1/3)$. 
The 1-forms $\alpha_{\perp,||\mu}$ 
transform under the $G$ as
\begin{eqnarray}
 \alpha_{\perp\mu}&\rightarrow& h(\pi,g_L,g_R)\cdot\alpha_{\perp\mu}\cdot h^\dagger(\pi,g_L,g_R)\,,\nonumber\\
 \alpha_{||\mu} &\rightarrow& h(\pi,g_L,g_R)\cdot\alpha_{||\mu}\cdot h^\dagger(\pi,g_L,g_R)- i \partial_\mu h(\pi,g_L,g_R)\cdot h^\dagger(\pi,g_L,g_R).
\end{eqnarray}

We include the vector meson field 
as a matter field (a la Callan-Coleman-Wess-Zumino)  
in the adjoint representation, $\rho_\mu=\rho_\mu^a T_a$, 
which transforms homogeneously 
under the chiral symmetry as 
$\rho_\mu\rightarrow h(\pi,g_R,g_L)\cdot\rho_\mu\cdot h^\dagger (\pi,g_L,g_R)$. 
The chiral-covariant derivative for the $\rho_\mu$ 
is then defined by 
$D_\mu\rho_\nu=\partial_\mu\rho_\nu-i[\alpha_{||\mu},\rho_\nu]$. 
The chiral invariant Lagrangian   
including the $\rho$ meson and the $\pi$ meson is 
thus written as 
\begin{eqnarray}
{\cal L}&=&F_\pi^2{\rm tr}[\alpha_{\perp\mu}\alpha^\mu_\perp]
+ \frac{F_\pi^2}{4} {\rm tr}[\hat{\chi}^\dag + \hat{\chi}] \nonumber\\
&&-\frac{1}{2}{\rm tr}[\rho_{\mu\nu}\rho^{\mu\nu}]+m_\rho^2{\rm tr}[\rho_\mu\rho^\mu]
-i{\cal G}{\rm tr}[\rho_{\mu\nu}\alpha_\perp^\mu\alpha_\perp^\nu],
\label{lag1form}
\end{eqnarray}
where 
$\rho_{\mu\nu}=D_\mu\rho_\nu-D_\nu\rho_\mu$ 
and ${\cal G}$ is a coupling constant related to the $\rho-\pi-\pi$ 
vertex. 
In Eq.(\ref{lag1form}) 
we also introduced the pion mass term (the second term in the first line) 
in which 
the chiral invariance is ensured by the 
spurion field $\hat\chi=\xi_L\chi\xi_R^\dagger$ 
having the transformation law 
$\chi\to g_L\cdot \chi \cdot g^\dagger_R$.
When the $\chi$ gets the vacuum expectation value, 
$\langle \chi \rangle 
=m_\pi^2\cdot {\bm 1}_{2\times2}$, 
the second term gives the pion mass.

Expanding the 1-forms in Eq.(\ref{para-perp}) 
in powers of the pion field, 
we extract portions relevant to the present study,  
\begin{eqnarray}
{\cal L}_{\rho^0,\pi^\pm}&=&D_\mu\pi^+D^\mu\pi^--m_\pi^2\pi^+\pi^-
-\frac{1}{2}(\partial_\mu\rho^{0}_\nu-\partial_\nu\rho^0_\mu)(\partial^\mu\rho^{0\nu}-\partial^\nu\rho^{0\mu})
+\frac{1}{2}m_\rho^2\rho^0_\mu\rho^{0\mu}\nonumber\\
&&+ig_{\rho\pi\pi}\rho^{0\mu}(\partial_\mu\pi^+\cdot\pi^--\partial_\mu\pi^-\cdot\pi^+),
\label{lagrhopi}
\end{eqnarray}
where we defined 
$D_\mu\pi^\pm=(\partial_\mu\mp ieA_\mu)\pi^\pm$ 
for the charged pions, $\pi^\pm \equiv (\pi^1 \mp i \pi^2)/\sqrt{2}$,  
and used the equation of motion for the rho field to 
get the on-shell $\rho-\pi-\pi$ coupling, 
$g_{\rho\pi\pi}={\cal G}m_\rho^2/(4F_\pi^2)$ 
in the last line, which is experimentally $\simeq 6$ in the vacuum. 
The covariantized-pion kinetic term ($|D_\mu \pi^\pm|^2$) in Eq.(\ref{lagrhopi}) 
provides the charged-pion propagator under the magnetic field $B$ (in $A_\mu$). 
It can be expressed by the Schwinger's proper-time procedure~\cite{Schwinger:1951nm}. 
In the present analysis, instead of the Schwinger form, we shall take 
a form expanded in terms of the Landau levels as done in Ref.~\cite{Ayala:2004dx}.

Let us consider the constant magnetic field $B$ oriented along 
the $z$-direction in the position-space time. 
The $\pi^\pm$ propagator, $G^B(x,y)$, then takes the form   
\begin{eqnarray}
G^B(x,y)=\Phi(x,y)\int\frac{d^4p}{(2\pi)^4}G^B(p)e^{-ip\cdot (x-y)}
\,, 
\label{pipro}
\end{eqnarray}
where
\begin{eqnarray}
\Phi(x,y)&=&\exp\left[ie\int_y^xdx'_\mu A^\mu(x')\right]\,,\nonumber\\
G^B(p)
&=&
2i\sum_{l=0}^\infty (-1)^lL_l\left(2\frac{p_\perp^2}{eB}\right) \exp\left[
\frac{p_\perp^2}{eB}
\right]
\frac{1}{p_{||}^2-(2l+1)eB-m^2_\pi}\,,
\label{modpi}
\end{eqnarray}
with $L_l\left(2\frac{p_\perp^2}{eB}\right)$ being the Laguerre polynomials 
labeling the Landau level as $l$.   
In Eq.(\ref{modpi}) we have introduced the following notations for 
the four-momenta:  
\begin{eqnarray}
&&p_{||}^\mu=(p^t,0,0,p^z),\,\,\,\,\,p_\perp^\mu=(0,p^x,p^y,0)
\,, \nonumber\\
&&p_{||}^2=(p^t)^2-(p^z)^2,\,\,\,p_\perp^2=-(p^x)^2-(p^y)^2  
\,.
\end{eqnarray}
Note also that the functional $\Phi(x,y)$ obviously depends on the gauge.    
Here we shall choose the gauge so as to set $\Phi \equiv 1$.

From the propagator expression in Eq.(\ref{modpi}), 
one should note 
that since nonzero constant magnetic field is present, 
no matter how small it is,   
the Lorentz invariance in four-dimension has been lost: 
the propagation of the $\pi^\pm$ is now confined to 
the z-direction parallel to the magnetic field.
This is the consequence of the dimensional reduction.

\section{rho meson propagator on the magnetic-dimensional reduction} 
\label{loopcprrection}

%%%%%%%%%%%%%%%%%%%%%%%%%%%%%%%%%%%%%%%%
%%%%%%%%%%%%%%%%%%%%%%%%%%%%%%%%%%%%%%%%

In this  section, 
based on the Lagrangian Eq.(\ref{lagrhopi}) 
and the $\pi^\pm$ propagator $G^B$ in Eq.(\ref{modpi}) 
(with $\Phi=1$), 
we shall compute and evaluate  
the $\pi^\pm$-loop corrections to the $\rho^0$ propagator 
at the one-loop level. 
As noted in the last paragraph of Sec.~\ref{model}, 
due to the magnetic-dimensional reduction 
the $\rho^0$ propagator and polarization structure 
no longer take the Lorentz-covariant form. 
To see the effect of the significant reduction,  
we shall hereafter take the LLL approximation and explicitly 
evaluate how the effect of the magnetic-dimensional reduction for the $\pi^\pm$ 
propagator is transferred to the $\rho^0$-polarization structure, 
mass and widths.

By including the one-loop correction arising from 
the $\rho-\pi-\pi$ vertex with the coupling strength in Eq.(\ref{lagrhopi}),  
the resultant inversed-$\rho^0$ propagator is expressed as 
\begin{eqnarray}
D^{-1}_{\mu\nu} 
=D^{-1}_{F\mu\nu} + 
\Pi_{\mu\nu}, 
\label{D-inv}
\end{eqnarray}
where $D^{-1}_{F\mu\nu}(p)=-(p^2-m^2_{\rho})g_{\mu\nu}+p_\mu p_\nu$ denotes
the free inversed-propagator and $\Pi_{\mu\nu}$ is the self-energy function,   
\begin{eqnarray}
i\Pi_{\mu\nu}=
-4(g_{\rho\pi\pi})^2\int\frac{d^4k}{(2\pi)^4}k_\mu k_\nu G^B(k-\frac{p}{2}) 
G^B(k+\frac{p}{2})
\,. \label{Pi:func}
\end{eqnarray} 
In the LLL approximation (with only $l=0$ kept in Eq.(\ref{modpi})),
the $\Pi_{\mu\nu}$ is decomposed by reflecting the dimensional reduction:  
\begin{eqnarray}
i\Pi_{\mu\nu}(p_{||}^2,p_\perp^2)=
i\left(\Pi_{||}^{S}(p^2_{||},p_\perp^2)g_{||\mu\nu}+
\Pi_{||}^{T}(p^2_{||},p_\perp^2)(p_{||}^2g_{||\mu\nu}-p_{||\mu}p_{||\nu})+
\Pi_{\perp}^{S}(p^2_{||},p_\perp^2)g_{\perp\mu\nu}\right)\,,
\label{loopfactor}
\end{eqnarray}
where $g_{||\mu\nu}={\rm diag}(1,0,0-1)$ and $g_{\perp \mu\nu}={\rm diag}(0,-1,-1,0)$, 
and  
\begin{eqnarray}
\Pi_{||}^{S}(p_{||}^2,p_{\perp}^2)& = & 
\frac{2(eB)}{(4\pi)^2}(g_{\rho\pi\pi})^2e^{\frac{p_\perp^2}{2eB}}
\int_0^1dx\Bigl(-2\ln\frac{\Lambda^2}{\Delta}+p_{||}^2(2x-1)^2\frac{1}{\Delta}\Bigl) 
\,, \nonumber\\
\Pi_{||}^{T}(p_{||}^2,p_{\perp}^2)&=& 
-\frac{2(eB)}{(4\pi)^2}(g_{\rho\pi\pi})^2e^{\frac{p_\perp^2}{2eB}}\int_0^1dx\frac{(2x-1)^2}{\Delta}\,, 
\nonumber\\
\Pi_{\perp}^{S}(p_{||}^2,p_{\perp}^2)&=& 
-\frac{2(eB)}{(4\pi)^2}(g_{\rho\pi\pi})^2e^{\frac{p_\perp^2}{2eB}}\int_0^1dx\frac{eB}{\Delta} 
\,, \label{loop-func}
\end{eqnarray} 
with $\Delta=x(x-1)p_{||}^2+m_\pi^2+eB$. 
(The relevant formulas for the Feynman parameter integrals in Eq.(\ref{loop-func}) 
are presented in Appendix~\ref{para-int}.) 
In evaluating the loop integrals, one has encountered  
the divergent term, which has been regularized by 
the dimensional regularization with the $D=2$ pole 
being replaced by the cutoff dependence $\Lambda$. 
Note that the overall factors of $(eB)$ in Eq.(\ref{loop-func}) 
come from the loop integrations of the pion momentum along the 
perpendicular direction ($k_\perp$ in Eq.(\ref{Pi:func})). 
That is the consequence of the dimensional reduction.

By performing the inversion of Eq.(\ref{D-inv}), 
one finds the propagator $D_{\mu\nu}$ in the presence of 
the magnetic field $(F_{\mu\nu}= \partial_\mu A_\nu - 
\partial_\nu A_\mu = B 
(\delta_{\mu 1} \delta_{\nu 2} - \delta_{\nu 1} \delta_{\mu 2})$), 
which can generically 
be decomposed into four independent polarization 
structures~\cite{Ritus:1972ky,Shabad:1975ik,Hattori:2012je}:      
\begin{eqnarray}  
D_{\mu\nu}=D_p\frac{p_\mu p_\nu}{p^2}+D_L\frac{L_\mu L_\nu}{L^2} +D_Q\frac{Q^*_\mu Q_\nu}{Q^*\cdot Q} +D_G\frac{G_\mu G_\nu}{G^2}
\,, 
\end{eqnarray} 
where the polarization vectors $L_\mu, Q_\mu$ and 
$G_\mu$ are defined as 
\begin{eqnarray} 
&&L^\mu=F^{\mu\nu}p_\nu=-B(0,p^y,-p^x,0)
\,, \nonumber\\
&&Q^\mu=\frac{i}{2}\epsilon^{\mu\nu\rho\sigma}F_{\rho\sigma}p_\nu
=iB(-p^z,0,0,-p^t)
\,, \nonumber\\
&&G^\mu=(p^2/L_\mu L^\mu)F^{\mu\nu}F_{\nu\lambda}p^\lambda+p^\mu=\left(p^t,-\frac{p_{||}^2}{p_\perp^2}p^x,-\frac{p_{||}^2}{p_\perp^2}p^y,p^z\right)
\,, \label{polar-1} 
\end{eqnarray}
and 
\begin{eqnarray}
D_p&=&
\frac{i}{m_\rho^2+\Pi_{||}^S\frac{p^2_{||}}{p^2}+\Pi_{\perp}^S
\frac{p^2_\perp}{p^2}}
\,, \nonumber\\
D_L&=&\frac{i}{m_\rho^2-p^2+\Pi_{\perp}^S}
\,, \nonumber\\
D_Q&=&\frac{i}{m_\rho^2-p^2+\Pi_{||}^S+p^2_{||}\Pi_{||}^T}
\,, \nonumber\\
D_G&=&\frac{i}{m_\rho^2-p^2+\Pi^S_{||}\frac{p_\perp^2}{p^2}
+\Pi_\perp^S\frac{p^2_{||}}{p^2}}
\,. \label{LQG}
\end{eqnarray} 
It turns out that 
the polarization mode along with the $D_p$ corresponds to 
the unphysical degree of freedom, like a scalar mode.  
(One can easily check it by constructing the equation of motion 
corresponding to the $D_p$ propagator form and finding that 
it satisfies the Klein-Gordon equation, so it is nothing but 
a scalar with spin $S=0$.)  
For the other three physical modes,  
$L_\mu,Q_\mu$ and $G_\mu$, 
the $\rho^0$ effective masses are defined as
\begin{eqnarray}
M_L^2&=&m_\rho^2+{\rm Re}\,\Pi_{\perp}^S(p^t=M_L,\vec{p}=\vec{0}) 
\,, \nonumber\\
M_Q^2&=&m_\rho^2+{\rm Re}\,\Pi_{||}^S(p^t=M_Q,\vec{p}=\vec{0})+M_Q^2{\rm Re}
\, \Pi_{||}^T(p^t=M_Q, \vec{p}=\vec{0}) 
\,, \nonumber\\
M_G^2&=&m_\rho^2+{\rm Re}\,
\Pi_\perp^S(p^t=M_G,\vec{p}=\vec{0}) 
\,, \label{masses}
\end{eqnarray}
where $\vec{p}=(p^x,p^y,p^z)$.  
To be canonical, the 
neutral 
$\rho^{(L)}$, $\rho^{(Q)}$ and $\rho^{(G)}$ fields are 
rescaled by the field renormalization constants 
$Z_{L}$, $Z_Q$ and $Z_G$ respectively:
\begin{eqnarray}
Z_L^{-1}&=&
1-\frac{\partial {\rm Re}\, \Pi_{\perp}^S(p^2_{||},p_\perp^2)}{\partial p^2_{||}}\Biggl|_{p^t=M_L,{\vec p}=\vec{0}} 
\,, \nonumber\\
Z_Q^{-1}&=& 
1-\frac{\partial\left( {\rm Re}\, \Pi_{||}^S(p^2_{||},p_\perp^2)+p^2_{||}{\rm Re}\,\Pi_{||}^T(p^2_{||},p_\perp^2)\right)}{\partial p^2_{||}}\Bigl|_{p^t=M_Q,{\vec p}=\vec{0}}
\,, \nonumber\\
Z_G^{-1}&=& 
1-\frac{\partial{\rm Re}\,\Pi_\perp^S(p^2_{||},p_\perp^2) }{\partial p_{||}^2}\Biggl|_{p^t=M_G,{\vec p}=\vec{0}} 
\,. 
\label{zfactor}
\end{eqnarray}
By expanding the propagators around the effective masses of 
$\rho^{(L)}$, $\rho^{(Q)}$ and $\rho^{(G)}$ fields,  
the functions $D_L,D_Q$ and $D_G$ in Eq.(\ref{LQG}) 
take the forms 
\begin{eqnarray} 
D_L&=& 
\frac{-i Z_L}{p^2_{||}-M_L^2
+Z_L\left(1-\frac{\partial {\rm Re}\, 
\Pi_{\perp}^S}{\partial p^2_{\perp}} 
\Bigl|_{p^t=M_L,\vec{p}=\vec{0}}
\right) p^2_{\perp}
-iZ_L{\rm Im}\, \Pi_{\perp}^S}+ 
\cdots 
\,, \nonumber\\
D_Q&=&
\frac{-iZ_Q}{
p^2_{||}-M_Q^2
+Z_Q\left(1-\frac{\partial\bigl( {\rm Re}\, \Pi_{||}^S+p^2_{||}{\rm Re}\, \Pi_{||}^T\bigl)}{\partial p^2_\perp}\Bigl|_{p^t=M_Q,\vec{p}=\vec{0}}\right) p^2_\perp
-iZ_Q{\rm Im}\, \Pi_{||}^S-iZ_Q{\rm Im}\, \Pi_{||}^T \cdot p^2_{||}
}+\cdots
\,, \nonumber\\
D_G&=&
\frac{-i Z_G}{p^2_{||}-M_G^2
+Z_G\left(1-\frac{\partial\bigl(\frac{p_\perp^2}{p^2}\Pi^S_{||}
+\frac{p_{||}^2}{p^2}\Pi_\perp^S \bigl)}{\partial p^2_\perp}
\Bigl|_{p^t=M_G,\vec{p}=\vec{0}} \right)p^2_\perp
-iZ_G{\rm Im}\, \Pi^S_{||} \cdot \frac{p_\perp^2}{p^2} 
-iZ_G{\rm Im}\, \Pi_\perp^S \cdot \frac{p_{||}^2}{p^2}} 
+ \cdots 
\, , \label{D-s}
\end{eqnarray} 
where the ellipses denote terms having no pole structure.

As to the decay width to the LLL-charged pions 
($\pi^\pm_{(l=0)}$), we assume the Breit-Wigner form for the 
propagators to 
extract the imaginary parts in Eq.(\ref{D-s}): 
\begin{eqnarray}
\Gamma(\rho^{(L)}\to\pi^+_{(l=0)}\pi^-_{(l=0)})&=&
-\frac{Z_L{\rm Im}\, \Pi^S_\perp(M_L,\vec{0})}{M_L} 
\,, \nonumber\\
\Gamma(\rho^{(Q)}\to\pi^+_{(l=0)}\pi^-_{(l=0)})&=&
-Z_Q\frac{{\rm Im}\, \Pi_{||}^S(M_Q,\vec{0})
+M^2_Q{\rm Im}\, \Pi_{||}^T(M_Q,\vec{0})}{M_Q}
\,, \nonumber\\
\Gamma(\rho^{(G)}\to\pi^+_{(l=0)}\pi^-_{(l=0)})&=&-\frac{
Z_G{\rm Im}\, \Pi_\perp^S(M_G, \vec{0})}{M_G}
\,. \label{widths}
\end{eqnarray}
Note from 
Eqs.(\ref{masses}) and (\ref{zfactor})  
that 
the mass and the decay width    
for the $\rho^{(L)}$ 
coincide with those of the $\rho^{(G)}$. 

%%%%%%%%%%%%%%%%%%%%%%%%%%%%%%%%%%%%%%%%
%%%%%%%%%%%%%%%%%%%%%%%%%%%%%%%%%%%%%%%%

\section{The magnetic dependence of masses and widths}  
\label{numerical}

In the previous section we derived the formulae for masses and 
widths at the one-loop level of the chiral effective model, 
which have been decomposed into the intrinsic 
polarization modes ($\rho^{(L)},\rho^{(Q)},\rho^{(G)}$) 
by the magnetic-dimensional reduction. 
Thus we are now ready to numerically study the magnetic dependence of 
the rho meson masses and widths.

First of all, to be fully consistent with 
the one-loop level computation, 
we expand the formulae of masses and widths in 
Eqs.(\ref{masses}) and (\ref{widths}),   
with the Feynman parameter integrals in 
Eq.(\ref{loop-func}) properly evaluated,   
up to terms of order of 
${\cal O}((g_{\rho\pi\pi}/4\pi)^2)$, 
to get  
\begin{eqnarray}
M_{L/G}^2&=&
\begin{cases}
m_\rho^2+ 
8(eB)^2 \left( \frac{g_{\rho\pi\pi}}{4\pi} \right)^2
\frac{1}{m_\rho\sqrt{m_\rho^2-4(m_\pi^2+eB)}}
\ln\frac{\sqrt{m_\rho^2-4(m_\pi^2+eB)}+m_\rho}{2\sqrt{m_\pi^2+eB}}\nonumber\\
\,\,\,\,\,\,\,\,\,\,\,\,\,\,\,\,\,\,\,\,\,\,\,\,\,\,\,\,\,\,\,\,\,\,\,\,\,\,\,\,\,\,\,\,\,\,\,\,\,\,\,\,\,\,\,\,\,\,\,\,\,\,\,\,\,\,\,\,\,\,\,\,\,\,\,\,\,\,\,\,\,\,\,\,\,\,\,\,\,\,\,\,\,\,\,\,\,\,\,\,\,\,\,\,\,\,\,\,\,\,\,\,\,\,\,\,\,\,\,\,\,\,\,\,\,\,\,\,\,\,\,\,\,\,\,
({\rm for}\,\,\,\,\,eB<\frac{m_\rho^2-4m_\pi^2}{4})
\\
m_\rho^2- 8(eB)^2 \left( \frac{g_{\rho\pi\pi}}{4\pi} \right)^2
\frac{1}{m_\rho\sqrt{4(m^2_\pi+eB)-m_\rho^2}}
\arctan\left(\frac{m_\rho}{\sqrt{4(m^2_\pi+eB)-m_\rho^2}}\right)
\nonumber\\
\,\,\,\,\,\,\,\,\,\,\,\,\,\,\,\,\,\,\,\,\,\,\,\,\,\,\,\,\,\,\,\,\,\,\,\,\,\,\,\,\,\,\,\,\,\,\,\,\,\,\,\,\,\,\,\,\,\,\,\,\,\,\,\,\,\,\,\,\,\,\,\,\,\,\,\,\,\,\,\,\,\,\,\,\,\,\,\,\,\,\,\,\,\,\,\,\,\,\,\,\,\,\,\,\,\,\,\,\,\,\,\,\,\,\,\,\,\,\,\,\,\,\,\,\,\,\,\,\,\,\,\,\,\,\,
({\rm for}\,\,\,\,\,\frac{m_\rho^2-4m_\pi^2}{4}<eB)\,,
\end{cases}
\\
M^2_{Q}&=&
\begin{cases}
m_\rho^2- 4(eB) \left( \frac{g_{\rho\pi\pi}}{4\pi} \right)^2
\Biggl(
\ln\frac{ \Lambda^2}{m^2_\pi+eB}+2-\frac{2}{m_\rho}\sqrt{m_\rho^2-4(m^2_\pi+eB)}
\ln\frac{\sqrt{m_\rho^2-4(m^2_\pi+eB)}+m_\rho}{2\sqrt{m_\pi^2+eB}}
\Biggl)
\nonumber\\
\,\,\,\,\,\,\,\,\,\,\,\,\,\,\,\,\,\,\,\,\,\,\,\,\,\,\,\,\,\,\,\,\,\,\,\,\,\,\,\,\,\,\,\,\,\,\,\,\,\,\,\,\,\,\,\,\,\,\,\,\,\,\,\,\,\,\,\,\,\,\,\,\,\,\,\,\,\,\,\,\,\,\,\,\,\,\,\,\,\,\,\,\,\,\,\,\,\,\,\,\,\,\,\,\,\,\,\,\,\,\,\,\,\,\,\,\,\,\,\,\,\,\,\,\,\,\,\,\,\,\,\,\,\,\,
({\rm for}\,\,\,\,\,eB<\frac{m_\rho^2-4m_\pi^2}{4})
\\
m_\rho^2- 4(eB) \left( \frac{g_{\rho\pi\pi}}{4\pi} \right)^2 \Biggl(
\ln \frac{\Lambda^2}{m^2_\pi+eB}+2-
\frac{2}{m_\rho}\sqrt{4(m^2_\pi+eB)-m_\rho^2}\arctan\frac{m_\rho}{\sqrt{4(m^2_\pi+eB)-m_\rho^2}}
\Biggl)
\nonumber\\
\,\,\,\,\,\,\,\,\,\,\,\,\,\,\,\,\,\,\,\,\,\,\,\,\,\,\,\,\,\,\,\,\,\,\,\,\,\,\,\,\,\,\,\,\,\,\,\,\,\,\,\,\,\,\,\,\,\,\,\,\,\,\,\,\,\,\,\,\,\,\,\,\,\,\,\,\,\,\,\,\,\,\,\,\,\,\,\,\,\,\,\,\,\,\,\,\,\,\,\,\,\,\,\,\,\,\,\,\,\,\,\,\,\,\,\,\,\,\,\,\,\,\,\,\,\,\,\,\,\,\,\,\,\,\,
({\rm for}\,\,\,\,\,\frac{m_\rho^2-4m_\pi^2}{4}<eB)\,,
\end{cases}\nonumber\\
\label{effmass}
\end{eqnarray}
and 
\begin{eqnarray}
\Gamma(\rho^{(L/G)}\to\pi^+_{(l=0)}\pi^-_{(l=0)})&=&
\frac{(eB)^2(g_{\rho\pi\pi})^2}{8\pi m_\rho^2}
\sqrt{\frac{4 }{m_\rho^2-4(m_\pi^2+eB)}}
\,\,\,\,\,\,\,\,\,
({\rm for}\,\,\,\,\,eB<\frac{m_\rho^2-4m_\pi^2}{4})\,,\nonumber\\
\Gamma(\rho^{(Q)}\to\pi^+_{(l=0)}\pi^-_{(l=0)})&=&
\frac{eB(g_{\rho\pi\pi})^2}{\pi m_\rho^2}
\sqrt{\frac{m_\rho^2-4(m_\pi^2+eB)}{4}}
\,\,\,\,\,\,\,
({\rm for}\,\,\,\,\,eB<\frac{m_\rho^2-4m_\pi^2}{4}).
\label{effwidth}
\end{eqnarray}
Here one should also notice from Eq.(\ref{effmass}) that  
the consistency of the one-loop computation 
including the constant magnetic scale requires 
the $eB$ to be constrained in such a way that 
the one-loop terms in total should be smaller than 
the vacuum term $(m_\rho^2)$. 
Thus the magnetic scale is bounded from above as 
\begin{equation} 
eB \ll  \frac{4\pi m_\rho^2} {g_{\rho\pi\pi} } 
\sim 1\, {\rm GeV}^2 
\,. \label{upper} 
\end{equation} 
Furthermore, since our computation has been restricted only to  
the LLL approximation, the magnetic scale $eB$ actually has 
the lower bound: 
it is set by requiring 
the 
$(eB)$ not to exceed the scale above which 
the decay channel involving at least one pion labeled as the the next-to-LLL 
is open, namely, 
\begin{equation} 
\Bigl(\sqrt{m_\pi^2+eB}+\sqrt{m_\pi^2+3eB}\Bigl)^2 > m_\rho^2
\,, \label{lower}
\end{equation} 
where the $(m_\pi^2 + 1(3) eB)$ 
correspond to the LLL (the next-to-LLL) pion mass.

\begin{figure}[htbp]
\begin{tabular}{cc}
 \begin{minipage}{0.5\hsize}
  \begin{center}
   \includegraphics[width=8cm]{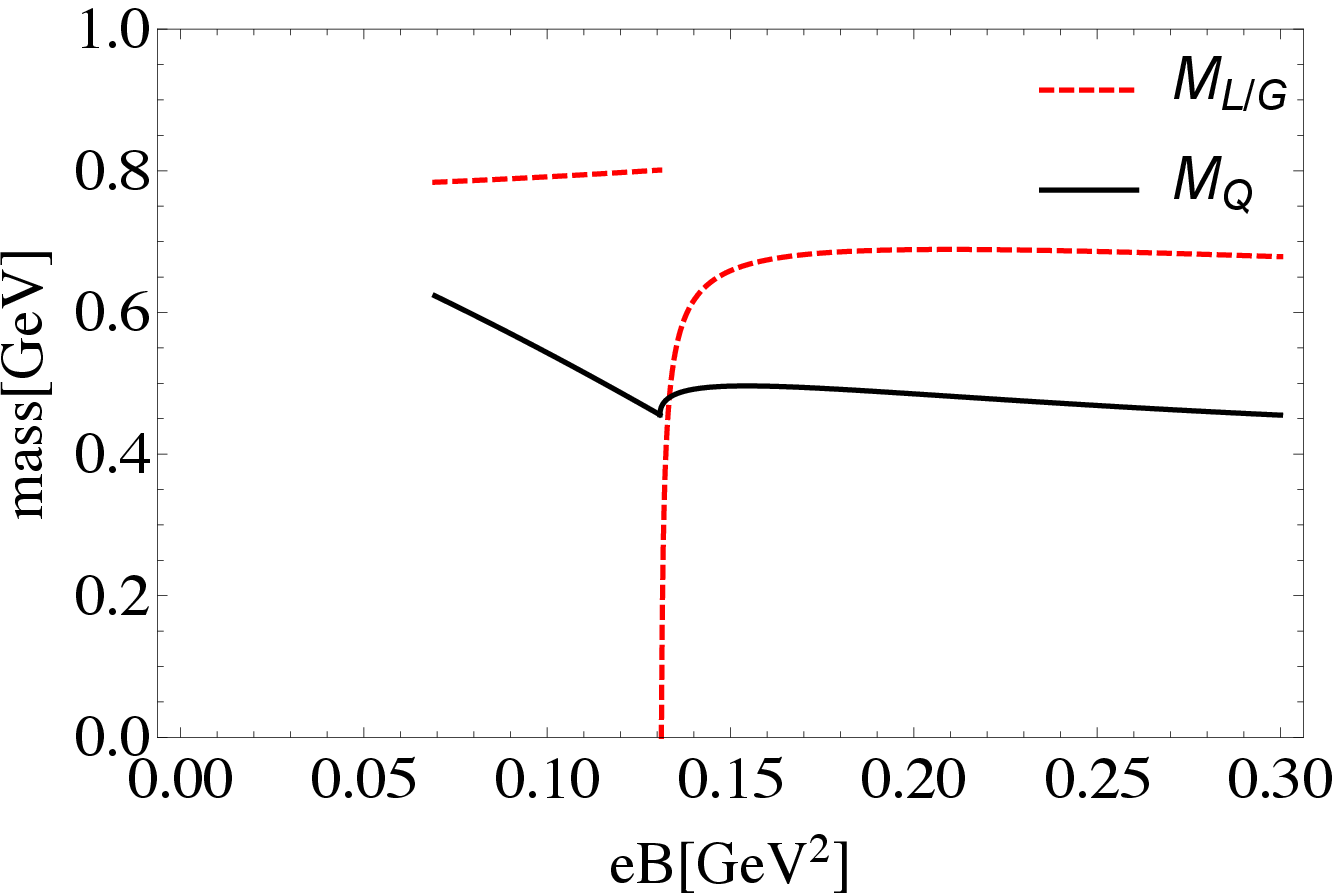}
  \end{center}
 \end{minipage} 
 \begin{minipage}{0.5\hsize}
  \begin{center}
   \includegraphics[width=8cm]{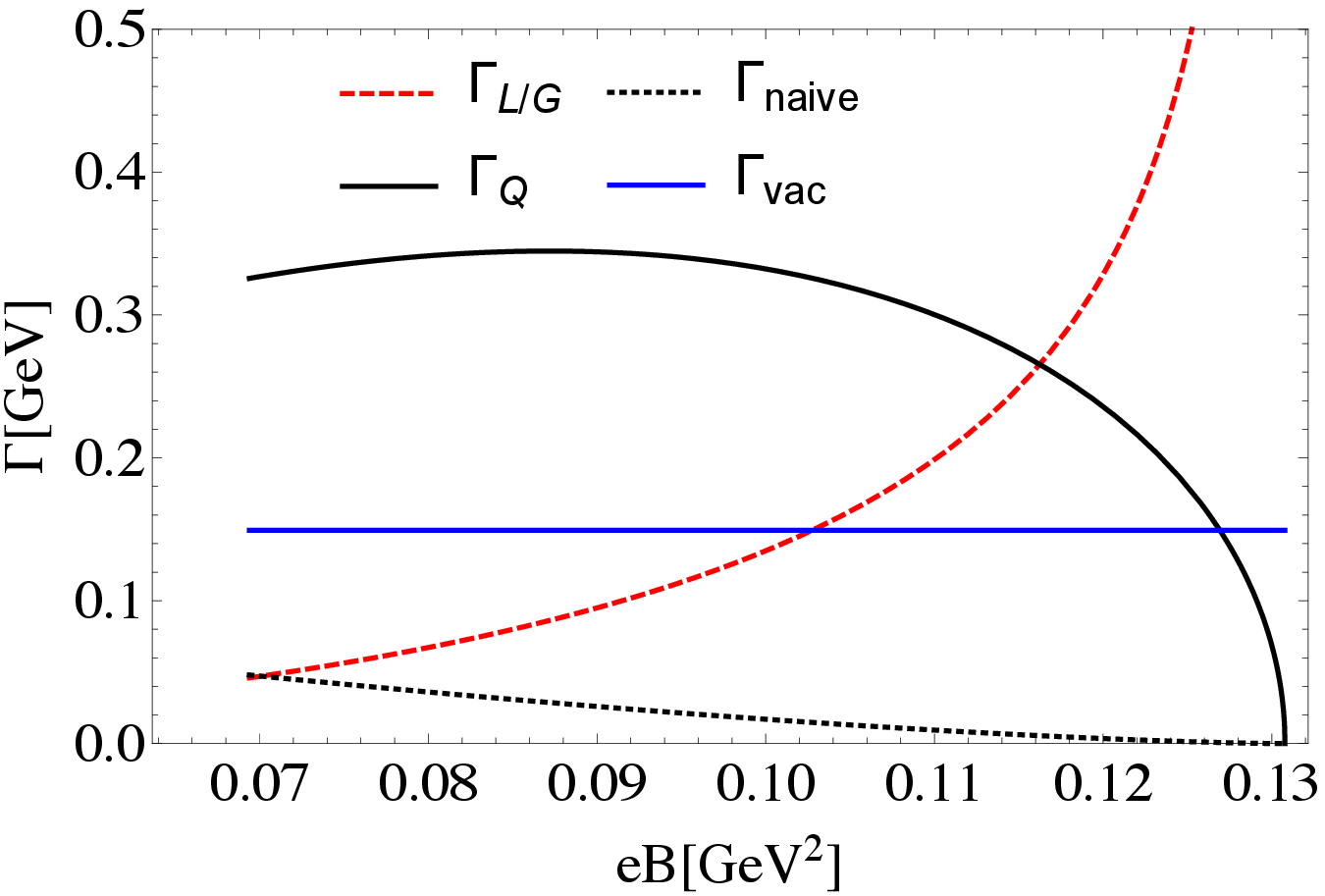}
  \end{center}
 \end{minipage}
 \end{tabular}
 \caption{ The magnetic-dependences of the rho masses (left panel) and 
widths (right panel) in the LLL approximation.  
In the right panel 
the naive expectation of the magnetic dependence 
on the decay width following from Eq.(\ref{width:naive}) 
has also been displayed together with 
the vacuum-width value estimated from Eq.(\ref{width:vac}).}  
\label{masswidth}
\end{figure}

In Fig.~\ref{masswidth} 
we plot the masses in Eq.(\ref{effmass}) and 
the decay widths in Eq.(\ref{effwidth}) as a function of 
the magnetic scale $(eB)$.  
We have used the experimental values in the vacuum~\cite{Agashe:2014kda},  
$m_{\pi^\pm}=0.140 \, {\rm GeV}$, 
$m_\rho=0.775\, {\rm GeV}$ and 
$g_{\rho\pi\pi}=5.98$ 
(which is estimated by the $\rho\to\pi\pi$ decay width in the 
vacuum), and taken the cutoff scale 
$\Lambda$ as $\Lambda \sim 4\pi F_\pi \simeq 1$ GeV. 
In the figure the magnetic scale has been constrained (from below) 
so as to satisfy the condition in Eq.(\ref{lower}), 
i.e., $eB \gtrsim 0.07\,{\rm GeV}^2$.

We first discuss the magnetic dependences of the decay widths of  
for $\rho^{(L/G)}$ and $\rho^{(Q)}$. As clearly seen in   
the right panel of Fig.~\ref{masswidth}, 
the magnetic dependence of the rho meson decay rate 
is more complicated than that expected from 
the naive observation (Eq.(\ref{width:naive})):  
as the $eB$ scale gets larger than 
the lowest $eB\simeq 0.07\, {\rm GeV}^2$, 
both the $\rho^{(L/G)}$ and $\rho^{(Q)}$ widths start to increase. 
As to the $\rho^{(Q)}$ width, 
it reaches a peak at around $eB \simeq 0.09 \,{\rm GeV}^2$ 
to go down after that, 
while the $\rho^{(L/G)}$ width develops the size monotonically. 
At the critical magnetic scale $eB_c=(m_\rho^2-4m_\pi^2)/4 \simeq 0.13\,{\rm GeV}^2$,  
the difference in magnetic dependences drastically gets prominent:  
the $\rho^{(L/G)}$ width diverges at the critical point, 
while the $\rho^{(Q)}$ width goes to zero.  
These tendencies can be understood from Eq.(\ref{effmass}): 
the decay width to the LLL pion, $\Gamma_{L/G}(\pi^+_{(l=0)}\pi^-_{(l=0)})$, develops like 
\begin{equation} 
\Gamma_{L/G}(\pi^+_{(l=0)} \pi^-_{(l=0)}) 
\sim 
\frac{(eB)^2} {\sqrt{m_\rho^2 -
 4(m_\pi^2 + eB)}} 
\,, \label{LG-scale}
\end{equation} 
which diverges at the critical point $eB_c$,  
while the $\Gamma_Q(\pi^+_{(l=0)} \pi^-_{(l=0)})$ scales like 
\begin{equation}
\Gamma_Q(\pi^+_{(l=0)}\pi^-_{(l=0)}) 
\sim (eB)\sqrt{m_\rho^2 -
 4(m_\pi^2 + eB)} 
\,, \label{Q-scale} 
\end{equation}  
which has the peak (at around $eB \simeq 0.09\,{\rm GeV}^2$).  
This discrepant tendency would actually imply the difference in the  
polarization (spin) sensitivity 
related to the Landau-level number conservation, 
as will be discussed later.

As to the magnetic dependence of  
the $\rho^{(L/G)}$ and $\rho^{(Q)}$ masses, 
actually, it looks somewhat complicated compared to the widths as depicted in   
the left panel of Fig.~\ref{masswidth}:    
below the critical magnetic scale $eB_c \simeq 0.13\,{\rm GeV}^2$, 
$M_{L/G}$ gets larger, while $M_Q$ becomes smaller 
as the $eB$ increases from the lowest value $\simeq 0.07\,{\rm GeV}^2$. 
This is thought to happen due to the net contributions coming from 
the complicated functional forms at around the lowest magnetic scale 
(see Eq.(\ref{effmass}). 
At the critical point $eB_c$, 
the discrepancy in magnetic dependences for $M_{L/G}$ and $M_{Q}$ 
drastically gets large: 
the $M_{L/G}$ becomes non-analytic at the critical point, 
while the $M_{Q}$ still looks continuous. 
The discontinuity for $M_{L/G}$ 
is actually related to the ill 
behavior for the $\rho_{L/G}$ decay width at the critical point 
(see Eq.(\ref{effwidth}) and the right panel of Fig.~\ref{masswidth}). 
Above the critical magnetic scale $eB_c \simeq 0.07\,{\rm GeV}^2$,   
the effect of the magnetic-dimensional reduction 
looks fairly mild for both $M_{L/G}$ and $M_Q$. 
(Actually, this tendency will be altered 
when one could ideally increase the magnetic field up to the scale above 
the upper limit $eB \sim 1 \, {\rm GeV}^2$ in Eq.(\ref{upper}): 
looking at Eq.(\ref{effmass}) one can easily see that the 
asymptotic behaviors of $M_{L/G}$ and $M_Q$, in 
the strong magnetic field limit, go like 
$M_{L/G}^2 \sim - eB$ and $M_Q^2 \sim + eB$.)

%%%%%%%%%%%%%%%%%%%%%%%%%%%%%%%%%%%%%%%%
%%%%%%%%%%%%%%%%%%%%%%%%%%%%%%%%%%%%%%%%
\section{Discussion} 
\label{dis}

In the previous section we have numerically evaluated 
the magnetic-dependences of the rho mass and decay width, 
which are classified with respect to the rho-meson polarization structures 
including the effect of the magnetic-dimensional reduction. 
In this section, we shall discuss possible interpretations for 
our findings from several views of the field theoretical ground.

\subsection{Moving onto the rest-frame representation}

We have so far analyzed the rho meson masses and decay rates 
by decomposing the polarization states defined in an active frame 
where the rho mesons are energetically moving in the three-space dimension 
(see Eq.(\ref{polar-1})). 
Instead, we shall here choose a rest frame, in which the rho meson has zero three-space 
momentum ($\vec{p}=\vec{0}$), to rephrase the results in the previous section, 
in terms of spin associated with the magnetic-direction, $s_z$. 
This would help us make comparison with other works based on the $s_z$-spin component.

To this end,  
we use the following polarization vectors irreducibly decomposed with respect to the $s_z$:  
\begin{eqnarray}
\epsilon_1^\mu&=&\frac{1}{\sqrt{2}}(0,1,i,0) \;\;\;\;\;\; (s_z=1,\;\;\;{\rm for}\;\rho^0)\,, 
\nonumber\\
\epsilon_2^\mu&=&\frac{1}{\sqrt{2}}(0,1,-i,0) \;\;\;(s_z=-1,\;\;\;{\rm for}\;\rho^0)\,, 
\nonumber\\
b^\mu&=&(0,0,0,1) \;\;\;\;\;\;\;\;\;\;\;(s_z=0,\;\;\;{\rm for}\;\rho^0)\,, 
\nonumber\\
u^\mu&=&(1,0,0,0)
\,. 
\label{spinpolarization}
\end{eqnarray}
The inversed-propagator of the rho meson is then decomposed into the form:  
\begin{eqnarray}
D^{-1}_{\mu\nu}(p^t,\vec{0})
&=&
m_\rho^2 g_{\mu\nu}-(p^t)^2 g_{\mu\nu}+
\left(p^t\right)^2u_\mu u_\nu
+
\Pi_{||}^S(p^{t},\vec{0})(u^\mu u^\nu - b^\mu b^\nu)
\nonumber \\ 
&& 
-\Pi_{||}^T(p^{t},\vec{0}) p_{t}^2b_\mu b_\nu
- \Pi_\perp^S(p^{t},\vec{0})(\epsilon^{*}_{1\mu}\epsilon_{1\nu}+\epsilon^{*}_{\mu2}\epsilon_{2\nu}) 
\,. 
\end{eqnarray}
The corresponding $\rho^0$ meson propagator can be expressed 
by using the polarization vectors in Eq.(\ref{spinpolarization}) to 
be 
\begin{eqnarray}
D_{\mu\nu}=D^{(s_z=+1)}\epsilon^{*\mu}_1\epsilon^\nu_1
+D^{(s_z=-1)}\epsilon^{*\mu}_2\epsilon^\nu_2 
+D^{(s_z=0)} b_\mu b_\nu+D^{(u)} u_\mu u_\nu 
\,. 
\end{eqnarray} 
Consequently, $D^{(s_z=+1)},D^{(s_z=-1)},D^{(s_z=0)}$ and $D^{(u)}$ are obtained 
as the functions of $\Pi$s as
\begin{eqnarray}
D^{(s_z=+1)}&=&\frac{i}{m_\rho^2-p^2_t+\Pi_\perp^S(p^{t},\vec{0})}
\,, \nonumber\\
D^{(s_z=-1)}&=&\frac{i}{m_\rho^2-p^2_t+\Pi_\perp^S(p^{t},\vec{0})} 
\,, \nonumber\\
D^{(s_z=0)}&=&\frac{i}{m_\rho^2-p^2_t+\Pi_{||}^S(p^{t},\vec{0})
+\Pi_{||}^T(p^{t},\vec{0})p_t^2} 
\,, \nonumber\\
D^{(u)}&=&\frac{i}{m_\rho^2+\Pi_{||}^S(p^{t},\vec{0})}
\, . \label{D-spin}
\end{eqnarray}
It is obvious from Eqs.(\ref{LQG}) and (\ref{D-spin}) 
that the masses and widths for the physical spin-1 modes $(D^{(s_z=\pm 1,0)})$ 
are respectively 
identical to those for the physical-active polarization modes ($D^{L,G,Q}$), 
namely,      
\begin{eqnarray}
&& M_{L/G} 
= M_{(s_z= \pm 1)}\, , \qquad 
M_{Q}=M_{(s_z=0)}\,, \nonumber \\ 
&& \Gamma(\rho^{(L/G)}\to\pi^+_{(l=0)}\pi^-_{(l=0)})
= \Gamma(\rho^{(s_z=\pm 1)}\to\pi^+_{(l=0)}\pi^-_{(l=0)}) 
\,, \nonumber\\  
&& \Gamma(\rho^{(Q)}\to\pi^+_{(l=0)}\pi^-_{(l=0)})
=\Gamma(\rho^{(s_z=0)}\to\pi^+_{(l=0)}\pi^-_{(l=0)}) 
\,. 
\end{eqnarray}
Thus, the results on the $\rho^{(L/G)}$ and $\rho^{(Q)}$ 
obtained in the previous section can be reinterpreted 
as those for the $\rho^{(s_z=\pm 1)}$ and $\rho^{(s_z=0)}$, respectively.

\subsection{
Polarization (spin)-dependence sensitive to Landau-level number-conservation?}  

In the previous section, Sec.~\ref{numerical}, 
we observed somewhat ill behaviors for the 
$\rho^{(L/G)}$ meson (i.e. $\rho^{(s_z=\pm 1)}$): 
the discontinuity for the mass and the divergence for the widths 
at the critical magnetic scale $eB_c \simeq 0.13\,{\rm GeV}^2$ 
(see Fig.~\ref{masswidth}).    
As to this result, here we shall address a possible interpretation 
in relation to the Landau-level number-conservation.

First of all, once encountering such an ill tendency,  
one might simply think that the wavefunction-renormalization factor 
($Z^{(L/G)}$) 
should diverge at the point (see the width formula in Eq.(\ref{widths})).  
In the present situation, however, 
it is not the case: we have explicitly checked that 
the $Z^{(L/G)}$ keeps finite values even at the critical magnetic scale.  
Therefore, the ill behavior seems to have nothing to do with the 
wavefunction-renormalization factor.

Now we would suspect that the issue could be related to 
the conservation of the Landau-level number:     
actually, if going beyond the LLL approximation in evaluating 
the pion-loop integral, 
one could find that terms involving pions carrying different Landau levels 
are present in the $(L/G)$-polarization ($s_z=\pm 1$) state of the rho meson. 
Note from Eq.(\ref{polar-1}) that 
only the $L/G$ polarization mode 
carries the spatial-momentum perpendicular ($\perp$) to the magnetic field, 
which contaminates the Landau-level number conservation
when couples to pions via the $\rho-\pi-\pi$ vertex like 
$\rho_\perp \pi^+_{(l_1)} \partial_\perp \pi^-_{(l_2)}$, 
resulting in the Landau-level number violation. 
Thus, the ill behavior about the $\rho^{(L/G)}$ (or $\rho^{(s_z=\pm 1)}$) 
may be understood by the disastrous breaking of the Landau-level 
number, which originates from the naive LLL truncation, 
hence the result on this polarization mode may be unreliable at this moment.

The well-defined magnetic-dependence of the $\rho^{(L/G)}$ 
would be obtained when one fully sums up the infinite tower of 
the Landau levels.

In contrast, 
the polarization mode of the $\rho^{(Q)}$ (or $\rho^{(s_z=0)}$), which   
accompanies only the momentum parallel to the magnetic field (see Eq.(\ref{polar-1})),     
is harmless against the Landau-level number conservation
because there does not exist a coupling form like  
$\rho_{||} \pi^+_{(l_1)} \partial_\perp \pi^-_{(l_2)}$  
breaking the Landau-level number when couples to pions.  
Thus, to this polarization mode, the LLL truncation is safely doable, 
hence we might have arrived at the well-defined magnetic-dependences  
as seen from Fig.~\ref{masswidth}.

Consequently, 
the result on the lifetime, the inverse of 
$\Gamma(\rho^{(Q)} (\rho^{(s_z=0)}) \to \pi^+_{(l=0)}\pi^-_{(l=0)})$ 
in Fig.~\ref{masswidth} and Eq.(\ref{effwidth}),  
is manifestly physical:  
recall the constrained region of the magnetic scale in Eq.(\ref{lower}) 
up to the critical scale ($eB_c\simeq 0.13\,{\rm GeV}^2$), 
$0.07 \lesssim eB \lesssim 0.13 \,{\rm GeV}^2$, 
which only allows the decay channel to the LLL-charged pion pair, 
namely, 
\begin{equation} 
\Gamma(\rho^{(Q)} (\rho^{(s_z=0)}) \to \pi^+_{(l=0)}\pi^-_{(l=0)})
= \Gamma(\rho^{(Q)} (\rho^{(s_z=0)}) \to \pi^+\pi^-) 
\,, \label{exact}
\end{equation} 
where $\pi^\pm$ include the full Landau-level tower.   
This clearly suggests a new ``fate" of the rho meson 
in the magnetic field, which would be more complicated 
than the naive expectation followed from Eq.(\ref{width:naive}), 
as displayed in the right panel of Fig.~\ref{masswidth}.

Unlike the case of the lifetime, 
the mass of the $\rho^{(Q)}$ ($\rho^{(s_z=0)}$) might 
get significant contributions from 
the higher Landau-levels even in the present 
restricted magnetic domain,      
$0.07 \lesssim eB \lesssim 0.13 \,{\rm GeV}^2$. 
One could compare the present result 
on the $\rho^{(s_z=0)}$ mass with other works 
in Refs.~\cite{Luschevskaya:2012xd,Andreichikov:2013zba,Luschevskaya:2014lga,Liu:2014uwa,Luschevskaya:2015bea,Andreichikov:2016ayj}.   
To make a conclusive answer to the validity of the 
LLL approximation for the mass estimation, 
one would need more rigorous argument including 
higher Landau levels.

The full computations regarding the $\rho^{(L/G)}$ mode including 
the infinite tower of the Landau levels would be also of importance, 
to be pursued in the future work.

%%%%%%%%%%%%%%%%%%%%%%%%%%%%%%%%%%%%%%%%
\section{Summary}
%%%%%%%%%%%%%%%%%%%%%%%%%%%%%%%%%%%%%%%%%%%%%
\label{summ}

In summary,  
we have attempted to access the naive 
expectation on the lifetime of the neutral rho meson 
in the magnetic field (Eq.(\ref{width:naive})), 
by explicitly computing 
the charged pion loop correction to the neutral rho meson 
propagator based on a chiral effective model. 
To see the significance of the magnetic-dimensional reduction-effect, 
we simply took the LLL approximation.  
We found that the magnetic field significantly 
separates the rho-meson polarization states 
in a nontrivial way, compared to the vacuum case, 
which is due to the magnetic-dimensional reduction 
(Eqs.(\ref{polar-1}) and (\ref{LQG})). 
According to the intrinsic polarization decomposition, 
the neutral rho mass and width are split as well, 
so the magnetic-dependences show up with respect to 
the polarization (spin) modes, labeled by $L/G$ and 
$Q$, or $s_z=\pm 1, 0$ for the physical modes.   
Then we numerically evaluated the magnetic-dependences 
of the masses and the widths respectively for the 
polarization (spin) modes. 
Of particular interest is that 
as the magnetic field increases,  
the rho width for the spin $s_z=0$ starts to develop, 
reach a peak, to be vanishing at the critical magnetic field 
to which the folklore refers (Fig.~\ref{masswidth}). 
This result is exact at the LLL approximation (Eq.(\ref{exact})), 
and would suggest that the life of the neutral rho meson 
in the magnetic field may be more complicated 
than the naive expectation.

A possible correlation between the 
spin-dependent magnetic scaling, the Landau-level 
number-conversation and the validity of the LLL approximation 
was also discussed.

\vspace*{20pt}
\noindent 
%%%%%%%%%%%%%%
{\it Note added}: 
After completion of the present manuscript, 
we noticed a paper (arXiv:1610.07887),  
in which a similar computation on the charged 
pion loop contribution to the neutral rho meson 
decay width has been made. 
In that paper, the spin-dependent decomposition 
due to the magnetic-dimensional reduction 
has not clearly been addressed, though, 
some portion of what we found in the present paper might be 
overlapped with theirs.

%%%%%%%%%%%%%%%%%%%%%%%%%%%%%%%%%%%%%%%%%%%%%
\acknowledgments
We would like to thank Kazunori Itakura for
enlightening discussions 
and Masayasu Harada and Hiroki Nishihara  for useful comments. 
This work was supported in part by 
the JSPS Grant-in-Aid for Young Scientists (B) \#15K17645 (S.M.).

\appendix
\section{The Feynman parameter integrals}
\label{para-int}

In this Appendix we present the Feynman parameter 
integrals relevant to the evaluation of the 
charged pion loop diagram in Sec.~\ref{loopcprrection}.

\begin{eqnarray}
\int_0^1dx\ln\frac{\Lambda^2}{\Delta}
=
\begin{cases}
\ln \Lambda^2+2-\ln\left(m^2_\pi+eB\right)-
\frac{2}{\sqrt{p^2_{||}}}\sqrt{4(m^2_\pi+eB)-p^2_{||}}\arctan\sqrt{\frac{p^2_{||}}{4m^2_\pi+eB-p^2_{||}}}\nonumber\\
\,\,\,\,\,\,\,\,\,\,\,\,\,\,\,\,\,\,\,\,\,\,\,\,\,\,\,\,\,\,\,\,\,\,\,\,\,\,\,\,\,\,\,\,\,\,\,\,\,\,\,\,\,\,\,\,\,\,\,\,\,\,\,\,\,\,\,\,\,\,\,\,\,\,\,\,\,\,\,\,\,\,\,\,\,\,\,\,\,\,\,\,\,\,\,\,\,\,\,\,\,\,\,\,\,\,\,\,\,\,\,\,\,\,\,\,\,\,\,\,\,\,\,\,\,\,\,\,\,\,\,\,\,\,\,
\left({\rm for}\,\,\,p_{||}^2<4(m_\pi^2+eB)\right)
\\
\\
\ln \Lambda^2+2-\ln\left(m^2_\pi+eB\right)-\frac{2}{\sqrt{p^2_{||}}}\sqrt{p^2_{||}-4(m^2_\pi+eB)}
\Biggl[\ln\frac{\sqrt{p^2_{||}-4(m^2_\pi+eB)}+\sqrt{p^2_{||}}}{2\sqrt{m_\pi^2+eB}}-i\pi\Biggl]\nonumber\\
\,\,\,\,\,\,\,\,\,\,\,\,\,\,\,\,\,\,\,\,\,\,\,\,\,\,\,\,\,\,\,\,\,\,\,\,\,\,\,\,\,\,\,\,\,\,\,\,\,\,\,\,\,\,\,\,\,\,\,\,\,\,\,\,\,\,\,\,\,\,\,\,\,\,\,\,\,\,\,\,\,\,\,\,\,\,\,\,\,\,\,\,\,\,\,\,\,\,\,\,\,\,\,\,\,\,\,\,\,\,\,\,\,\,\,\,\,\,\,\,\,\,\,\,\,\,\,\,\,\,\,\,\,\,\,
\left({\rm for}\,\,\,4(m_\pi^2+eB)<p^2_{||}\right)\,,
\end{cases}\nonumber\\
\end{eqnarray}
\\
\begin{eqnarray}
\int_0^1dx\frac{1}{\Delta}
=
\begin{cases}
\frac{4}{\sqrt{p^2_{||}}\sqrt{4(m^2_\pi+eB)-p^2_{||}}}
\arctan\left(\frac{p^2_{||}}{\sqrt{p^2_{||}}\sqrt{4(m^2_\pi+eB)-p^2_{||}}}\right)
\,\,\,\,\,\,\,\,\,\,\,\,\,\,\,
\left({\rm for}\,\,\,p_{||}^2<4(m_\pi^2+eB)\right)
\\
\\
\frac{1}{\sqrt{p^2_{||}}\sqrt{p^2_{||}-4(m_\pi^2+eB)}}
\Biggl[
-4\ln\frac{\sqrt{p^2_{||}-4(m_\pi^2+eB)}+\sqrt{p^2_{||}}}{2\sqrt{m_\pi^2+eB}}+2i\pi
\Biggl]
\,\,\,\,\,\,\,\,\,\,\,\,\,\,\,
\left({\rm for}\,\,\,4(m_\pi^2+eB)<p_{||}^2\right)\,,
\end{cases}
\end{eqnarray}
where $\Delta=x(x-1)p_{||}^2+m_\pi^2+eB$.

%%%%%%%%%%%%%%%%%%%%%%%%%%%%%%%%%%%%%%%%%%%%%
%%%%%%%%%%%%%%%%%%%%%%%%%%%%%%%%%%%%%%%%%%%%%

\end{document}